\documentclass {iopart}

\begin{document}
\jl{2}

\title{Calculation of hyperfine structure constants for ytterbium}

\author{S G Porsev\footnote[1]{E-mail: porsev@thd.pnpi.spb.ru},
Yu G Rakhlina and M G Kozlov}
\address{Petersburg Nuclear Physics Institute, \\
         Gatchina, St.-Petersburg district, 188350, RUSSIA}
\date{\today}
\submitted
%\maketitle
%------------------------------------------------------------------

\begin{abstract}
We calculate energies and hyperfine structure constants $A$ and $B$ for
low-lying levels of $^{173}$Yb. The latter is treated as a two-electron
atom.  To account for valence-valence correlations the configuration
interaction method is used. Valence-core correlations are accounted
for within the many-body perturbation theory which is used to construct
effective two-electron operators in the valence space.

\end{abstract}
%------------------------------------------------------------------

%======================
\section {Introduction}
%======================

Two years ago a method for calculation of the low-lying energy levels of
many-electron atoms was proposed \cite{DFK}. Calculations for Tl
\cite{DFK}, Ca, Sr, Ba and Yb \cite{KP97} and Ba \cite{DJ} confirmed its
effectiveness. Later this method was generalized for other observables,
such as hyperfine structure (HFS) constants \cite{DKPF} and
polarizabilities \cite{KP98}. Within this method valence-valence correlations
are treated non-perturbatively, while core-valence and core-core
correlations are treated within the many-body perturbation theory (MBPT).

Here we report calculations for the HFS constants of the low-lying levels
of ytterbium. Our interest to this atom is caused in part by an
extremely large parity nonconserving (PNC) amplitude for the
$^1S_0\,(6s^2) \rightarrow {}^3D_1\,(5d6s)$ transition. It was first
suggested by DeMille \cite{DeMille} and later confirmed in \cite{PRK,Das},
that this amplitude is approximately 100 times larger than PNC amplitude
for the $6s \rightarrow 7s$ transition in Cs and 10 times larger than that
for the $6p \rightarrow 7p$ transition in Tl. This makes Yb a very
promising candidate for the future PNC experiment. Therefore, an accurate
atomic calculations for Yb are needed.  Moreover, it is important to
have a reliable estimate of the accuracy of such calculations. PNC
amplitude is very sensitive to the behavior of the wave function in the
vicinity of the nucleus. Of course, the same applies to HFS constants.
That makes HFS calculations very important for the future analysis of the
theoretical accuracy for the PNC amplitude.

We also have a more general interest in ytterbium as an atom with the
relatively shallow $4f$ core shell. For such an atom it is not
obvious at all, that core-valence correlations can be treated by means of
the MBPT. In the earlier paper \cite{KP97} we have shown that the method
works for the low-lying energy levels and here we extend our analysis to
the HFS constants. We also show here, that with some caution it is possible
to calculate HFS constants of the levels, which lie above the core
excitation threshold.

In the next section of the paper we define effective operators for the
valence electrons and briefly describe how MBPT can be used to calculate
these operators (for more details see \cite{DFK,DKPF}). In
section~\ref{sec_HFS} we give some details of the calculations followed by
the discussion of the results in section~\ref{sec_Disc}.

%===================================================
\section {Effective operators for valence electrons}
%===================================================
\label{sec_Eff}

At present there are several methods of calculations for many-electron
atoms. For atoms with one electron above a closed core a conventional MBPT
is quite effective (see, for example, \cite{LM}). For atoms with several
valence electrons the multi configurational Hartree-Fock (MCHF) method is
often used \cite{GQ}. Lately the coupled-claster method became very popular
\cite{BJS,IK,IKI}. All these methods were used to calculate HFS constants
of different atoms (see, for example, \cite{L1,DFSS,ORFT,BD,JF,MP}).

The most complicated problem in precise atomic calculations is
associated with the necessity to treat valence-valence correlations,
core-valence, and core-core correlations. The former are usually too
strong to be treated perturbatively, while other types of correlations can
not be effectively treated within non-perturbative techniques, such as
MCHF or CI method.

Therefore, it is natural to try to combine MBPT with one of the
non-perturbative methods. In \cite{DFK} it was suggested to use MBPT in
order to construct an effective Hamiltonian for valence electrons. After
that, Schr\"odinger equation for valence electrons is solved within CI
approach. That allows to find the low-lying energy levels.
%(excitation energy should be smaller than that of the core).
In order to calculate other atomic observables, one need to construct
corresponding effective operators for valence electrons \cite{DKPF}. The
latter effectively account for the core-valence and core-core correlation,
which are of particular importance for such singular operators as HFS
ones.

Suppose, that we are interested in atomic levels with energies $E_i-E_0 <
\varepsilon$, where $E_0$ is the ground state energy. In the first
approximation we can assume that inner electrons with the
Hartree-Fock energies $|\varepsilon_n| \gg \varepsilon$ form the core,
which is described by the wave function
%- - - - - - - - - - - - - - - - - - - - - - - - - - - - - - - - -
\begin{eqnarray}
     &&\Psi_{\rm core} = (N_c !)^{-1/2}
     {\rm det}(\phi_1,\phi_2,\dots \phi_{N_c}),
\label{2.1}\\
     &&h_{\rm DF} \phi_n = \varepsilon_n \phi_n,
\label{2.2}
\end{eqnarray}
%- - - - - - - - - - - - - - - - - - - - - - - - - - - - - - - - -
where $h_{\rm DF}$ is the Dirac-Fock operator, and $N_c$ is the number of
core electrons.  Note, that this operator can include the field of some
valence electrons as well.  For example, it is natural to consider Yb as a
two-electron atom with the core $[1s^2 \dots 4f^{14}]$, and operator
$h_{\rm DF}$ can be calculated for configuration $1s^2 \dots 4f^{14}6s^2$
(so called, $V^N$-approximation, N is the number of electrons in the atom).

Let us say that many-electron wave function $\Psi$ belongs to the valence
subspace if core electrons are in the state (\ref{2.1}). We will denote
projector operator on this subspace by $P$. Then, operator $Q=1-P$ will
project on the subspace for which at least one of the core electrons is
excited to one of the states $\phi_n$ with $n > N_c$.

One can show \cite{DFK}, that Schr\"odinger equation
%- - - - - - - - - - - - - - - - - - - - - - - - - - - - - - - - -
\begin{eqnarray}
      H \Psi = E \Psi
\label{2.3}
\end{eqnarray}
%- - - - - - - - - - - - - - - - - - - - - - - - - - - - - - - - -
is equivalent to the following equation in the $P$-subspace for the
function $\Phi = P \Psi$:
%- - - - - - - - - - - - - - - - - - - - - - - - - - - - - - - - -
\begin{eqnarray}
     &&\left( PHP+ \Sigma(E) \right) \Phi =  E \Phi,
\label{2.4} \\
     &&\Sigma(E) = P V^\prime R_Q(E) V^\prime P,
\label{2.5}
\end{eqnarray}
%- - - - - - - - - - - - - - - - - - - - - - - - - - - - - - - - -
where $V^\prime$ is the operator of the residual Coulomb interaction and
$R_Q(E)$ is the Green's function in $Q$-subspace:
%- - - - - - - - - - - - - - - - - - - - - - - - - - - - - - - - -
\begin{eqnarray}
     &&V^\prime = H-H_0,
\label{2.6} \\
     &&R_Q(E) = Q \frac{1}{E-QHQ} Q.
\label{2.7}
\end{eqnarray}
%- - - - - - - - - - - - - - - - - - - - - - - - - - - - - - - - -
Operator $H_0$ is defined in terms of one-electron
operator (\ref{2.2}):
%- - - - - - - - - - - - - - - - - - - - - - - - - - - - - - - - -
\begin{eqnarray}
     H_0= \sum_{i=1}^{N} h_{\rm DF}(\vec{r}_i) - W.
\label{2.12}
\end{eqnarray}
%- - - - - - - - - - - - - - - - - - - - - - - - - - - - - - - - -
The constant $W$ in the right hand side of this equation is introduced to
compensate the double counting of the two-electron interaction in the
sum. It can be fixed, for example, by the requirement:
%- - - - - - - - - - - - - - - - - - - - - - - - - - - - - - - - -
\begin{eqnarray}
     \langle \Psi_{\rm core} |H_0| \Psi_{\rm core} \rangle =
     \langle \Psi_{\rm core} |H| \Psi_{\rm core} \rangle,
\label{2.12a}
\end{eqnarray}
%- - - - - - - - - - - - - - - - - - - - - - - - - - - - - - - - -
which implies, that
%- - - - - - - - - - - - - - - - - - - - - - - - - - - - - - - - -
\begin{eqnarray}
     W = W_0 \equiv
     \sum_{i=1}^{N_c} \varepsilon_i
     - \langle \Psi_{\rm core} |H| \Psi_{\rm core} \rangle.
\label{2.12b}
\end{eqnarray}
%- - - - - - - - - - - - - - - - - - - - - - - - - - - - - - - - -
It is also possible to use $W$ as a free parameter to match the energy
spectrum. This subject is discussed in more details elsewhere
\cite{KP1_98}.

Equations (\ref{2.3}) --- (\ref{2.7}) yield:
%- - - - - - - - - - - - - - - - - - - - - - - - - - - - - - - - -
\begin{eqnarray}
    &&\Psi =\left( P + R_Q(E) V^\prime P \right) \Phi,
\label{2.8}
\end{eqnarray}
%- - - - - - - - - - - - - - - - - - - - - - - - - - - - - - - - -
The orthonormality condition
$\langle \Psi_i|\Psi_k \rangle = \delta_{i,k}$
is approximately equivalent to the following condition for $\Phi$:
%- - - - - - - - - - - - - - - - - - - - - - - - - - - - - - - - -
\begin{eqnarray}
     && \langle \Phi_i |1 - \partial_E \Sigma(\bar{E})
     | \Phi_k \rangle \approx \delta_{i,k},
\label{2.9}
\end{eqnarray}
%- - - - - - - - - - - - - - - - - - - - - - - - - - - - - - - - -
where $\bar{E} \approx (E_i+E_k)/2$. Note, that only the last of equations
(\ref{2.4}) --- (\ref{2.9}) is approximate.

The operator in the left hand side of equation \eref{2.4} plays the role
of an effective Hamiltonian $H_{\rm eff}$. Equations \eref{2.4} ---
\eref{2.7} allow to construct it with the help of conventional
diagrammatic technique \cite{DFK,KP97}, which is based on the following
representation of the Green's function:
%- - - - - - - - - - - - - - - - - - - - - - - - - - - - - - - - -
\begin{eqnarray}
     &&R_Q(E) = R_Q^0(E) + R_Q^0(E) V^\prime R_Q(E),
\label{2.10} \\
     &&R_Q^0(E) = Q \frac{1}{E-QH_0 Q} Q.
\label{2.11}
\end{eqnarray}
%- - - - - - - - - - - - - - - - - - - - - - - - - - - - - - - - -
In particular, if we restrict ourselves to the second order MBPT, we have
to replace $R_Q(E)$ in the right hand side of \eref{2.5} by $R_Q^0(E)$.

Suppose now, that we know solutions of \Eref{2.4} and we want to use them
to calculate an observable $a$, which is associated with one-electron
operator $A$:
%- - - - - - - - - - - - - - - - - - - - - - - - - - - - - - - - -
\begin{eqnarray}
     &&a = \langle \Psi | A | \Psi \rangle.
\label{3.1}
\end{eqnarray}
%- - - - - - - - - - - - - - - - - - - - - - - - - - - - - - - - -
Let us define effective operator $A_{\rm eff}$ so, that
%- - - - - - - - - - - - - - - - - - - - - - - - - - - - - - - - -
\begin{eqnarray}
     &&a = \langle \Phi | A_{\rm eff} | \Phi \rangle.
\label{3.2}
\end{eqnarray}
%- - - - - - - - - - - - - - - - - - - - - - - - - - - - - - - - -
The following approximation for $A_{\rm eff}$ was suggested in
\cite{DKPF}
%- - - - - - - - - - - - - - - - - - - - - - - - - - - - - - - - -
\begin{eqnarray}
     &&A_{\rm eff} \approx PAP + PV^\prime R_Q^0(E)A_{\rm rpa}P
     + PA_{\rm rpa}R_Q^0(E)V^\prime P,
\label{3.4}
\end{eqnarray}
%- - - - - - - - - - - - - - - - - - - - - - - - - - - - - - - - -
where $A_{\rm rpa}$ corresponds to the well-known random-phase
approximation (RPA).

Note, that only the first order MBPT corrections are completely included
in (\ref{3.4}). Some second order corrections, the, so-called, structural
radiation, as well as many higher-order corrections, are omitted.
Nevertheless, this approximation accounts for several most important
MBPT corrections to all orders. Diagrammatic representation of \Eref{3.4}
can be found in \cite{DKPF}.

%==========================================
\section {Hyperfine structure calculations}
%==========================================
\label{sec_HFS}

We use basis representation for diagrams for effective Hamiltonian and for
effective HFS operators. That implies that all sums over intermediate
states run over some finite set of one-electron states. The latter
should effectively account for both discrete and continuous part of the
spectrum of Dirac-Fock operator \eref{2.2}.

In this calculation basis set included 167 four-component orbitals:
$1-23s$, $2-22p$, $3-22d$, $4-15f$, $5-15g$ and $6-13h$. Core orbitals as
well as valence orbitals $6s,7s$, $6p,7p$, and $5d,6d$ were obtained by
solving Dirac-Fock equations for appropriate configurations, while higher
virtual orbitals were formed, as described in
\cite{Bogdan,Bogdan1,DFK,DKPF}. After that operator \eref{2.2} was
diagonalized in order to have quasi-Dirack-Fock basis set.

This basis set was used to solve RPA equations for HFS operators and to
calculate diagrams for effective Hamiltonian and effective HFS operators.
After that, the solution of \Eref{2.4} was found in the CI approximation.
On this stage it was possible to truncate our basis set to $7-15s$,
$6-15p$, $5-14d$, $5-10f$ and $5-7g$ orbitals. We made complete CI on
this shorter basis set.

We calculated effective Hamiltonian $H_{\rm eff}$ in the second order of
MBPT. That means, that zero order approximation \eref{2.11} for the
Green's function was used in \eref{2.5}. The choice of the constant $W$
suggested by \Eref{2.12b} resulted in some underestimation of the binding
energy for levels of the configuration $5d6s$ and for $^1P_1^o$ level of
the configuration $6s6p$.  For these levels $W=W_0+0.4$~au was used.  Our
final results for the ground state are given in \tref{tab1} and for
several low-lying levels in \tref{tab2}.

On the next stage we used corresponding wave functions to calculate HFS
constants. These calculations are similar to those of the papers
\cite{DKPF} and \cite{KP98}. Results are given in tables~\ref{tab3}
and~\ref{tab4}.  In analysis of the theoretical accuracy, which is
done in the next section, it is very important to know the scale of
different contributions to the final answer.  Therefore, in
tables~\ref{tab3} and~\ref{tab4} we give results of the Dirac-Fock
calculation, of the conventional two-electron CI and of the two-electron CI
with effective Hamiltonian. Final values include effective operator
corrections \eref{3.4} and normalization correction \eref{2.9}. The latter
usually decreases the answer by 1\%--2\%.

%====================
\section {Discussion}
%====================
\label{sec_Disc}

We have pointed out above, that the effective Hamiltonian can be safely used
only for the energy levels below the core excitation threshold. For Yb
this threshold lies at 23189~cm$^{-1}$ above the ground state and many of
the levels from the \tref{tab2} lie higher. Nevertheless, theoretical
spectrum is quite good up to the level $^1$P$_1^o(6s7p)$, which
appears 1634~cm$^{-1}$ below its experimental position. This huge
discrepancy can be attributed to interaction with the $J=1$ level at
38422~cm$^{-1}$, which is supposed to belong to the configuration
$4f^{13}5d^26s$ \cite{Martin}. This level can also interact with
$^3$P$_1^o(6s7p)$, what makes calculations for this level less reliable.
For other levels from the \tref{tab2} there are no close levels with the
same $J$ and the same parity, which correspond to the excitations from the
$4f$-shell.

The perturber levels, discussed above, correspond to the poles of the
Green's function $R_Q(E)$ and, thus, to the poles of the operator
$\Sigma(E)$.  As long as we use approximate Green's function \eref{2.11}
instead of the exact one, the poles of $\Sigma(E)$ are shifted from the
physical poles. In order to use effective Hamiltonian, we have to keep
far enough from the poles of the exact Green's function and from the poles
of the approximate Green's function.  As we go further up above the core
excitation threshold,
%(e.g, above the levels obtained by 4f-electron excitations),
both sets of poles become denser. For this reason, it is
hardly possible to use effective operator technique there.

Let us now proceed to the calculations of HFS constants presented in the
tables~\ref{tab3} and~\ref{tab4}. In these tables the first two rows
correspond to the non-perturbative part of the calculation, while two lower
rows include MPBT corrections associated with the effective Hamiltonian and
the effective HFS operators. One can see, that average MBPT contribution is
about 30\% for magnetic constant $A$ and about 40\% for electric quadrupole
constant $B$. Thus, theoretical accuracy for the constant $A$ is normally
higher. Indeed, we have neglected higher order corrections to the effective
Hamiltonian and to the effective HFS operators, that can be justified only
if MBPT corrections are small.

For 6 levels
($^3$D$_2\,(5d6s)$,     $^3$D$_3\,(5d6s)$,
 $^3$S$_1\,(6s7s)$,     $^3$D$_1\,(6s6d)$,
 $^3$D$_3\,(6s6d)$, and $^3$P$_2^o\,(6s7p)$)
MBPT corrections to the constant $A$ are less than 25\%. For all these
levels the difference between the theory and experiment is within 3\%.
For the constant $B$ the smallest MBPT corrections (about 35\%) correspond
to levels $^3$D$_1\,(6s6d)$ and $^3$P$_1^o\,(6s6p)$, where we have
agreement with the experiment within 2\%. Constant $B$ for
$^3$S$_1\,(6s7s)$- state differs from zero only because of the
configuration interaction.

On the other hand, when MBPT corrections are 40\% or more,
the accuracy of the theory becomes uncertain. The most striking
discrepancy with the experiment takes place for levels
$^1$D$_2\,(5d6s)$,
$^3$D$_2\,(6s6d)$ and
$^1$P$_1^o\,(6s6p)$.
For all of them MBPT corrections appear to be 40\%--50\%.

The only exception of this rule is the level $^3$P$_1^o\,(6s7p)$, where
MBPT corrections for both constants are about 30\%, but even the sign of
the constant $B$ differs from the experiment.  As we pointed out above,
this level can interact with the perturber level $4f^{13}5d^26s (J=1)$,
which is located at 38422~cm$^{-1}$. Constant $B$ for this level is
relatively small and even small admixture of the perturber can result in
the change of its sign, while the larger constant $A$ should be less
affected by such a mixing.

%====================
\section {Conclusion}
%====================
\label{sec_Conc}

In this paper we have checked applicability of the CI+MBPT method for
ytterbium. The first core excitation here lies only 23189~cm$^{-1}$ above
the ground state, which seems to restrict us to only few lowest levels.
Actually, the method works quite well up to 38930~cm$^{-1}$,
where the first significantly perturbed level arise.

When the method is used for calculations of HFS constants, the accuracy of
calculations depends on the scale of MBPT corrections. When total MBPT
correction contributes less than 25\% to the value of the constant, the
accuracy appears to be better than 3\%. On the contrary, when MBPT
corrections account for more than 40\% of the theoretical value, the
accuracy becomes uncertain. In this case one can use MBPT correction only
as a rough estimate of the theoretical error. We expect, that similar
relation holds for other one-electron operators as well.

For the studied levels of Yb the average MBPT correction is approximately
30\% for the constant $A$ and 40\% for the constant $B$. Therefore,
precision calculations of the constant $A$ are possible for the majority of
levels, while precision calculations of the constant $B$ are possible only
as exception.

\ack

This work was supported in part by Russian Foundation for Basic Research,
Grant No~98-02-17663. SP is grateful to St.~Petersburg government for the
financial support. MK was supported by Engineering and Physical Sciencies
Research Council.

%====================
\section*{References}
%====================
\label{sec_Ref}

%==================================================================
\newpage
\begin{table}
\caption{Two-electron energies $E_{\rm val}$ of the Yb
ground state in different approximations (au). MBPT corrections were
calculated with $W=W_0$ (see \Eref{2.12b}).}

\label{tab1}
\begin{indented}
\lineup
\item[]\begin{tabular}{lcccc}
\br
DF$^{\rm a}$ & MBPT$^{\rm b}$ & CI$^{\rm c}$
& CI+MBPT$^{\rm d}$ & Experiment$^{\rm e}$\\
\mr
0.606810  &  0.654308 &  0.632398 &  0.677601 &  0.677584 \\
\br
\end{tabular}
\item[]$^{\rm a}${Single-configuration approximation.}
\item[]$^{\rm b}${Single-configuration approximation with MBPT
    correlation.}
\item[]$^{\rm c}${Conventional CI method for two outer electrons.}
\item[]$^{\rm d}${CI+MBPT method: two-electron CI with effective Hamiltonian
formed within second order of MBPT.}
\item[]$^{\rm e}${This is the sum of the first two ionization
potentials of Yb \cite{Martin}.}

\end{indented}
\end{table}

%############################################################
\begin{table}
\caption{The low-lying levels of Yb in $V^{N}$ approximation. $E_{\rm
val}$ is the two-electron energy and $\Delta$ is the interval from the
ground state. The multiplet splittings are given in parentheses. Levels
marked with {\dag} were calculated with $W=W_0+0.4$~au and the levels
marked with {\ddag} were calculated with $W=W_0+0.1$~au.}

\label{tab2}
\begin{indented}
\lineup
\item[]\begin{tabular}{llll}
\br
 & \multicolumn{2}{c}{Theory}
 & \multicolumn{1}{c}{Exper.$^{\rm a}$}\\
 Level  &$E_{\rm val}$ (au) & $\Delta$ (cm$^{-1}$)
 &$\Delta$ (cm$^{-1}$) \\
\mr
\multicolumn{4}{c}{Even states} \\
\mr
$^3$D$_1~(5d6s)$\dag&  -0.566086 &  24441        &    24489        \\
$^3$D$_2~(5d6s)$\dag&  -0.564919 &  24697 (256)  &    24752 (263)  \\
$^3$D$_3~(5d6s)$\dag&  -0.562415 &  25247 (550)  &    25271 (519)  \\
$^1$D$_2~(5d6s)$\dag&  -0.551500 &  27642        &    27678        \\
$^3$S$_1~(6s7s)$    &  -0.529514 &  32501        &    32695        \\
$^1$S$_0~(6s7s)$    &  -0.521725 &  34210        &    34351        \\
$^3$D$_1~(6d6s)$    &  -0.496053 &  39845        &    39809        \\
$^3$D$_2~(6d6s)$    &  -0.495902 &  39877 (33)   &    39838 (29)   \\
$^3$D$_3~(6d6s)$    &  -0.495384 &  39991 (114)  &    39966 (128)  \\
$^1$D$_2~(6d6s)$    &  -0.494733 &  40135        &    40062        \\
\mr
\multicolumn{4}{c}{Odd states} \\
\mr
$^3$P$_0^o~(6s6p)$\ddag &  -0.598858 &  17282        &    17288        \\
$^3$P$_1^o~(6s6p)$\ddag &  -0.595599 &  17997 (715)  &    17992 (704)  \\
$^3$P$_2^o~(6s6p)$\ddag &  -0.587613 &  19750 (1753) &    19710 (1718) \\
$^1$P$_1^o~(6s6p)$\dag&  -0.563354 &  25074        &    25068        \\
$^3$P$_0^o~(6s7p)$    &  -0.504463 &  38000        &    38091        \\
$^3$P$_1^o~(6s7p)$    &  -0.504096 &  38080 (80)   &    38174 (83)   \\
$^3$P$_2^o~(6s7p)$    &  -0.502388 &  38455 (375)  &    38552 (378)  \\
$^1$P$_1^o~(6s7p)$    &  -0.500224 &  38930        &    40564        \\
\br
\end{tabular}
\item[]$^{\rm a}${Ref. \cite{Martin}}
\end{indented}
\end{table}

%############################################################
%\newpage
\begin{table}
\caption{Magnetic dipole ($A$) and electric quadrupole ($B$) hyperfine
structure constants of low-lying even-parity levels for $^{173}$Yb.
The electric quadrupole moment is taken to be 2.80~b.}

\label{tab3}

\footnotesize
\lineup
\begin{tabular}{@{}lccccccccc}
\br
 & \multicolumn{1}{c}{$^3D_1$}
 & \multicolumn{1}{c}{$^3D_2$}
 & \multicolumn{1}{c}{$^3D_3$}
 & \multicolumn{1}{c}{$^1D_2$}
 & \multicolumn{1}{c}{$^3S_1$}
 & \multicolumn{1}{c}{$^3D_1$}
 & \multicolumn{1}{c}{$^3D_2$}
 & \multicolumn{1}{c}{$^3D_3$}
 & \multicolumn{1}{c}{$^1D_2$} \\

 & \multicolumn{1}{c}{$(5d6s)$}
 & \multicolumn{1}{c}{$(5d6s)$}
 & \multicolumn{1}{c}{$(5d6s)$}
 & \multicolumn{1}{c}{$(5d6s)$}
 & \multicolumn{1}{c}{$(6s7s)$}
 & \multicolumn{1}{c}{$(6s6d)$}
 & \multicolumn{1}{c}{$(6s6d)$}
 & \multicolumn{1}{c}{$(6s6d)$}
 & \multicolumn{1}{c}{$(6s6d)$} \\
\mr
\multicolumn{10}{c}{$A$ (MHz)}  \\
\mr
DF
& 447 & $-$200 & $-$348 & $-$18 & $-$1225 & 490 & $-$206 & $-$337 & 28  \\
CI
& 443 & $-$288 & $-$348 &    62 & $-$1489 & 633 & $-$468 & $-$437 & 238 \\
$H_{\rm eff}$
& 588 & $-$398 & $-$469 &   105 & $-$1910 & 830 & $-$677 & $-$567 & 380 \\
Total
& 596 & $-$351 & $-$420 &   131 & $-$1938 & 838 & $-$683 & $-$569 & 392 \\
Exper.$^{\rm a}$
 & 563(1)      & $-$362(2)   & $-$430(1) & 100(18) & $-$1879(10)
 &  818.7(4)   & $-$732.5(4) &$-$559.9(5)  & 438.5(4) \\
\mr
\multicolumn{10}{c}{$B$ (MHz)}  \\
\mr
DF
& 151 & 206 & 353 & 379 &  0   &  32 & 45 & 76 & 81  \\
CI
& 156 & 229 & 368 & 676 &  0.3 &  38 & 55 & 92 & 115 \\
$H_{\rm eff}$
& 219 & 314 & 492 & 693 &  0.2 &  38 & 62 & 92 & 115 \\
Total
& 290 & 440 & 728 &1086 &  0.2 &  58 & 96 &150 & 184 \\
Exper.$^{\rm a}$
 & 335(1)  & 482(22)
 & 909(29) & 1115(89)
 & $<$3(18) & 59.3(2)
 & 52.5(2) & 139.6(3)
 & 142.2(2) \\
\br
\end{tabular}
$^{\rm a}$ {HFS constants for configuration $5d6s$ were measured in
\cite{Toepper}, for $^3$S$_1$-state in \cite{Ber}, and for
configuration $6s6d$ in \cite{Jin}}

\end{table}

%###################################################################
%\newpage
\begin{table}
\caption{Magnetic dipole ($A$) and electric quadrupole ($B$) hyperfine
structure constants of low-lying odd-parity levels for $^{173}$Yb.}

\label{tab4}

\begin{indented}
\lineup
\item[]\begin{tabular}{@{}lccccc}
\br
& \multicolumn{1}{c}{$^3P_1^o$}
& \multicolumn{1}{c}{$^3P_2^o$}
& \multicolumn{1}{c}{$^1P_1^o$}
& \multicolumn{1}{c}{$^3P_1^o$}
& \multicolumn{1}{c}{$^3P_2^o$} \\
%& \multicolumn{1}{c}{$^1P_1^o$} \\

& \multicolumn{1}{c}{$(6s6p)$}
& \multicolumn{1}{c}{$(6s6p)$}
& \multicolumn{1}{c}{$(6s6p)$}
& \multicolumn{1}{c}{$(6s7p)$}
& \multicolumn{1}{c}{$(6s7p)$} \\
%& \multicolumn{1}{c}{$(6s7p)$} \\
\mr
\multicolumn{6}{c}{$A$ (MHz)}  \\
\mr
DF
& $-$664 & $-$527 & $-$ 25 & $-$667 & $-$507 \\      % & 128  \\
CI
& $-$765 & $-$556 &     98 &$-$1044 & $-$666 \\      % & 346  \\
$H_{\rm eff}$
& $-$1075& $-$741 &    187 &$-$1468 & $-$862 \\      % & 567  \\
Total
& $-$1094& $-$745 &    191 &$-$1488 & $-$871 \\      % & 574  \\
Exper.
& $-$1094.0(7)$^{\rm a}$    & $-$738$^{\rm b}$
&     59$^{\rm b}$          & $-$1144(1)$^{\rm c}$
& $-$854(2)$^{\rm c}$       \\
& $-$1094.2(6)$^{\rm d}$ \\
\mr
\multicolumn{6}{c}{$B$ (MHz)}  \\
\mr
DF
& $-$454 &  715 & 811 & $-$90 & 150  \\    %  & 165  \\
CI
& $-$533 &  860 & 428 & $-$80 & 163  \\     % & 195  \\
$H_{\rm eff}$
& $-$633 & 1013 & 642 & $-$82 & 192  \\     % & 219  \\
Total
& $-$822 & 1335 & 848 & $-$111& 264  \\     % & 285  \\
Exper.
 & $-$ 826.5(1)$^{\rm a}$ &   1312$^{\rm b}$
 &    605$^{\rm b}$  &     12(4)$^{\rm c}$
 &    267(25)$^{\rm c}$  \\
 & $-$827.2(5)$^{\rm d}$ \\
\br
\end{tabular}
\item[]$^{\rm a}$ {Ref. \cite{Toepper}};
$^{\rm b}$ {Ref. \cite{Jin}};
$^{\rm c}$ {Ref. \cite{Clark}};
$^{\rm d}$ {Ref. \cite{Budick}}
\end{indented}
\end{table}

\end{document}